\documentclass[twocolumn,superscriptaddress, prb]{revtex4-1}
\usepackage{amsfonts}
\usepackage{amssymb}
\usepackage{amsmath}
\usepackage{amsthm}
\usepackage{dsfont}

\usepackage{stackrel}
\usepackage{color}

\usepackage[pdftex]{graphicx,hyperref}

\usepackage{hyperref}

\newcommand{\avg}[1]{\left< #1 \right>}
\newcommand{\bra}[1]{\langle#1|}
\newcommand{\ket}[1]{|#1\rangle}

\newcommand{\tr}{\textrm{Tr}}

\begin{document}

\title{Eigenstate Thermalization and Representative States on Subsystems}

\author{Vedika Khemani}
\affiliation{Department of Physics, Princeton University, Princeton, NJ 08544}

\author{Anushya Chandran}
\affiliation{Perimeter Institute for Theoretical Physics, 31 Caroline Street N, Waterloo, Ontario, Canada N2L 2Y5}

\author{Hyungwon Kim}
\affiliation{Department of Physics, Princeton University, Princeton, NJ 08544}

\author{S. L. Sondhi}
\affiliation{Department of Physics, Princeton University, Princeton, NJ 08544}

\date{\today}

\begin{abstract}
We consider a quantum system $A  \cup B$ made up of degrees of freedom that can be partitioned into spatially disjoint regions
$A$ and $B$. When the full system is in a pure state in which regions $A$ and $B$ are entangled, the quantum mechanics of region $A$
described without reference to its complement is traditionally assumed to require a reduced density matrix on $A$. While this is certainly
true as an exact matter, we argue that under many interesting circumstances expectation values of typical operators {\it anywhere} inside $A$ can be computed from a suitable {\it pure} state on $A$ alone, with a controlled error.
We use insights from quantum  statistical mechanics---specifically the eigenstate thermalization hypothesis (ETH)---to argue for the
existence of such ``representative states''.

\end{abstract}

\maketitle


\section{Introduction}

In this paper we consider the following problem. Let $|AB \rangle$ be
a pure state of the quantum system $A  \cup B$ made up of degrees of freedom that can be partitioned into spatially disjoint regions $A$ and $B$ with $A$ being the smaller subregion.
We wish to find a pure state on region $A$, $\ket{\psi_A}$, which we can use for
practical purposes to reproduce expectation values of typical operators of
interest in region $A$. We will call such a state a ``representative state'' (RS)
on $A$.

Evidently, proceeding axiomatically would require us to define which operators
are ``typically of interest'' and what error is acceptable for ``practical
purposes''. With these defined we can then ask for what states $|AB \rangle$ and bipartitions $A$ and $B$ such RS can be found. We will not try
to carry out such an exercise in the abstract. Instead we will use ideas from
quantum statistical mechanics, notably the equivalence of ensembles and the
eigenstate thermalization hypothesis (ETH)\cite{Deutsch:1991ss,Srednicki:1994dw,Rigol:2008bh} to discuss several broad classes of
states for which one can usefully define RS. Possibly future work can fold our
concrete examples into a more general account.

The striking feature of a RS description of subsystems is that it dispenses with the entanglement between the degrees of freedom in $A$ and those outside.
This entanglement is at the root of the exact description by means of the
reduced density matrix
$$\rho_A = {\rm Tr}_B |AB \rangle \langle AB |$$
which is the textbook prescription for describing a subsystem. We are interested in replacing this exact description with an RS description.

The intuition for why it may be possible to replace $\rho_A$ with a single state on $A$ comes from writing $\rho_A$ in the suggestive form\cite{Li:2008pb} $$\rho_A = e^{-H_E}$$ which defines the entanglement Hamiltonian $H_E$ on $A$. In this form, $\rho_A$ is the canonical density matrix of $H_E$ at entanglement temperature $T_E = 1$,  and all physical observables in $A$ are derived from this ensemble: $\langle O_A \rangle_{T_E = 1} = \mbox{Tr}(\rho_A O_A)= \mbox{Tr}(e^{-H_E}O_A)$. If $H_E$ is assumed to be ``generic'' -- in the sense that we can do quantum statistical mechanics with it -- we can replace canonical averages with a single quantum state via the ETH. More concretely, the ETH assumes that eigenstate expectation values (EEVs) of  few-body  observables computed from individual eigenstates in an energy window match canonical or microcanonical averages in the thermodynamic limit. It follows that if $H_E$ satisfies the ETH, we can replace the canonical ensemble of $H_E$  with eigenstates of $H_E$ drawn from the right entanglement energy window. These states are the desired ``representative states''. Further, in cases where $H_E$ doesn't satisfy the ETH (e.g. $H_E$ is integrable/free or many-body localized\cite{Oganesyan:2007aa}), RS can be found for a smaller, more restricted class of observables in a manner to be discussed later. [We note that in a previous paper\cite{Chandran:2013qy} we have employed this strategy of doing statistical mechanics with $H_E$ to study the limits of the universality of the low-energy entanglement spectrum.]

In this article we will discuss three families of quantum states for which
an RS description can be provided. These are a) ground states of local
quantum Hamiltonians, b) highly excited states (those with a finite energy
density) of local Hamiltonians, and c) randomly picked states in Hilbert
space. For (a) and (b) we will consider subsystems $A$ such that both $A$ and $B$ are simply connected domains, while
for (c) we will consider arbitrary subsystems of $A  \cup B$. In all three
cases we use the number of spins/qubits in $A$, denoted by $|A|$,  as our control parameter
with the implicit ordering $1 \ll |A| \le | B|$. In this limit we
will argue that we can reproduce the expectation values of few-body
operators\footnote{One question we leave open is the meaning of ``few'',
where the corresponding question regarding ETH is still open.} on A to controlled accuracy by means of RS.

In detail, we start with a free fermion system for which $H_E$ is known to be free (and hence integrable)\cite{Peschel:2003aa}. While this is a ``non-generic'' case which doesn't permit us to use the full machinery of ETH, it nonetheless provides a transparent illustration of our ideas for a special class of operators that are ``orthogonal'' to the conserved quantities. We consider RS descriptions of both the ground state and highly-excited states of the free-fermion system.
We then generalize our results to ground and excited states of generic gapped, local quantum Hamiltonians. In this case, we provide evidence that $H_E$ will also be generic and we can use the ETH to argue for RS. Finally, we consider randomly picked vectors in Hilbert space where the RS can be obtained quite directly.  We conclude with some comments on generalizations and open questions.

\section{Free fermions}

We begin with a gapped free fermion model in 2D which illustrates the ideas and errors involved in a representative states description. Consider the dimerized hopping model in 2D:
\begin{align}
H = - \sum_{i,j} t^x_{i,i+1} \; c_{i,j}^{\dagger} c_{i+1,j} + t^y \; c_{i,j}^{\dagger} c_{i, j+1} +  h.c.
\end{align}
where $c_{i,j}$ are fermionic operators on sites $(i,j)$ of a 2D square lattice, the hopping in the $x$ direction, $t^x_{i,i+1}$, alternates between $1\pm\delta$, and $t^y$ is the hopping in the $y$ direction.
The Hamiltonian is readily diagonalized in momentum space, and there are two bands with momenta in the reduced Brillouin zone. At half filling, the model is gapped for either $t^y < \delta < 1$ or $\delta>1$ and $t^y <1$.

The entanglement Hamiltonian for free fermion systems is itself quadratic\cite{Peschel:2003aa}:
\begin{align}
 \rho_A = \frac{1}{Z} e^{-H_E}, \;\;\;\;\;\; H_E = \sum_{i=1}^{|A|} \epsilon_i f_i^{\dagger}f_i
 \label{eq:HE-FF}
 \end{align}
 where the operators $f_i$ live in $A$ and are related to the original fermionic operators by a canonical transformation, and $Z = \mbox{Tr} \rho_A$. The single-particle entanglement energies $\{ \epsilon_i \}$ are easily calculated through their monotonic relation with the eigenvalues $\xi_i$ of the correlation matrix $C_{\mathbf{r}\mathbf{r'}} \equiv \langle c_\mathbf{r}^{\dagger} c_\mathbf{r'} \rangle$ restricted to region $A$:
 \begin{align}
 \epsilon_i = \log \left(\frac{1-\xi_i}{\xi_i}\right).
 \end{align}
 Evidently, $H_E$ is also integrable, with the set of conserved quantities
 $f_i^{\dagger}f_i$.

We will show that we can find representative states in $A$ that reproduce canonical averages computed using $\rho_A$. However, the RS cannot be used to reproduce all few-body observables in $A$. Since $H_E$ is integrable (and thus non-generic for the purposes of the ETH), we must restrict ourselves to few-body observables that are roughly uniformly ``spread" over all conserved quantities in $H_E$. As our underlying Hamiltonian is translationally
invariant, we expect that momentum conservation is broken in $H_E$ by boundary effects alone so that the $f_i^{\dagger}f_i$ have a fair degree
of locality in momentum space. This indicates that operators which are local
in real space are good candidates for an RS description and we study these below.

We do this in turn for the system at zero and finite temperatures.

\subsection{$T=0$}

\begin{figure*}[htbp]
\includegraphics[width=\textwidth]{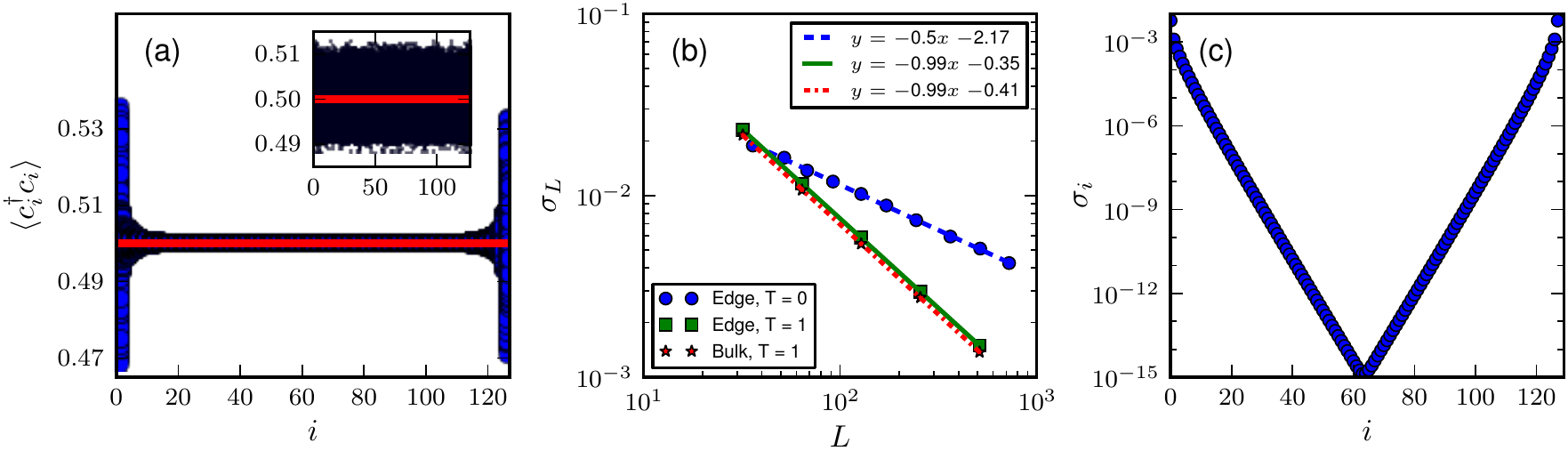}
\caption{ (a) ${}\bra{\psi_A}c^{\dagger}_i c_i\ket{\psi_A} $ plotted against the position $i$ for 100,000 randomly picked representative states  $\ket{\psi_A}$  in a dimerized free-fermion system of linear dimension $L = 256, L_A = 128$ and temperature $T =0$. The red line denotes the canonical average. The error is maximum for boundary operators. Inset: Same results for a system at temperature $T = 1$. In this case there is no discernible difference in the variance between boundary and bulk operators consistent with the volume law. (b)  Standard deviation of $\bra{\psi_A}c^{\dagger}_i c_i\ket{\psi_A} $ for $i$ either at the boundary or deep in the bulk plotted against system size at temperatures $T = 0, 1$. The plots confirm the $\sqrt{\frac{1}{L^{d-1}}}$ scaling of the error for boundary operators at $T = 0$, and the  $\sqrt{\frac{1}{L^{d}}}$ scaling for both boundary and bulk operators at finite $T$. (c) standard deviation of  $\bra{\psi_A}c^{\dagger}_i c_i\ket{\psi_A} $ as a function of position $i$,  showing exponential decay with distance from the boundary.    }
\label{Fig:FF_GS}
\end{figure*}

Pick a set of parameters $t^y$ and $\delta$ such that the Hamiltonian $H$ is gapped at half filling. At zero temperature, the system is in the ground state of $H$ on $A \cup B$. We trace over half the system with the entanglement cut along the $y$ axis to obtain $\rho_A$ and $H_E$ in the usual fashion. Gapped ground states are believed to satisfy an area law for the entanglement entropy\cite{Eisert:2010ab}:
$$S_E = -\mbox{Tr} \;\rho_A \log \rho_A \sim s L_A^{d-1}$$
where $L_A$ is the linear size of region $A$ and $d$ is the spatial dimenson.
In $d=1$, a rigorous proof of the above scaling
exists\cite{Hastings:2007kx,Arad_Improved}. The entanglement entropy is the thermal entropy of $H_E$ at $T_E = 1$; as this scales only with the area of the boundary, $H_E$ is morally a $(d-1)$-dimensional Hamiltonian whose low-energy excitations live on the boundary between $A$ and $B$.

The many-body eigenstates of $H_E$ are Slater determinants in terms of the $f$ operators in \eqref{eq:HE-FF}. For spatially local observables, the canonical ensemble of $H_E$ at $T_E = 1$ can be replaced by individual eigenstates:
we pick representative states $|\psi_A\rangle$ by filling single particle states $ f^{\dagger}_i|0\rangle$ with the Fermi-Dirac (FD) probability distribution at $T_E = 1$ and $\mu_E = 0$. Thus the representative states lie in an energy window that scales as $\sqrt{L_A^{d-1}}$ about the mean entanglement energy $\avg{H_E}_{T_E = 1} $.

Drawing states using the FD distribution ensures that averages for operators $\hat{\mathcal{A}}$ computed using the ensemble of RS agree with the canonical average of $H_E$. However, there are fluctuations from eigenstate to eigenstate within the energy window which can be shown to scale as
\begin{equation}
\avg{\hat{\mathcal{A}}}_{T_E = 1} = \bra{\psi_A}  \hat{\mathcal{A}} \ket{\psi_A} + O\left(\sqrt{\frac{1}{L_A^{d-1}}}\right).
\label{Eq:ErrFFCantoeig}
\end{equation}
The scaling follows from the expansion of the $\hat{\mathcal{A}}$ in the mode occupation basis: $\hat{\mathcal{A}}  =  \frac{1}{L_A^{d-1}}\sum_i \hat{n}_i  a(i) $, where $\hat{n}_i = f_i^\dagger f_i$ and $a(i)$ is a smooth function of the mode index $i$. In each RS, $\hat{n}_i =0,1$, while the probability that $\hat{n}_i = 1$ is given by the FD distribution. Further, the occupation numbers of different modes in the RS ensemble are uncorrelated. Thus, Eq.~\eqref{Eq:ErrFFCantoeig} follows from the central limit theorem. Observe that the fluctuations go to zero in the infinite volume limit for $d>1$.

We now present numerical evidence supporting our claims. For simplicity, we study expectation values of local density operators $\hat{\mathcal{A}}_{i} = c^{\dagger}_{i,0} c_{i,0}$, though more complicated $m$-local operators could also be considered. Note that translation invariance is preserved along the $y$ direction so operators are only labeled by $i$, their position along the $x$ axis.
The main plot in Fig.~\ref{Fig:FF_GS}(a) shows $\bra{\psi_A}\hat{\mathcal{A}}_i \ket{\psi_A} $ for 100,000 representative states $\ket{\psi_A}$ randomly picked with FD probabilities.
We work in a system of length $L = 256$ and $L_A = 128$, and consider $ \hat{\mathcal{A}}_{i} $ for all sites $i$ along the $x$ axis.
The red line is the canonical average $\langle \hat{\mathcal{A}}_i \rangle_{T_E = 1} = \mbox{Tr} \rho_A \hat{\mathcal{A}}_i$.
We see that the EEVs in representative states $\bra{\psi_A}\hat{\mathcal{A}}_i \ket{\psi_A} $ follow the canonical average $\langle \hat{\mathcal{A}}_i \rangle_{T_E = 1}$ quite closely, with the error being maximum for operators near the boundaries of $A$.
This is consistent with the picture that the $O(L^{d-1})$ eigenstates of $H_E$ that contribute to canonical averages resemble the starting ground state in the bulk of $A$ and only differ on the boundary.
Fig.~\ref{Fig:FF_GS}(b) (blue circles) shows the standard deviation of $\bra{\psi_A}\hat{\mathcal{A}}_i \ket{\psi_A} $ for $i$ at the boundary of $A$ for various system sizes confirming the $\sqrt{\frac{1}{L^{d-1}}}$ scaling of the error posited in \eqref{Eq:ErrFFCantoeig}. Finally, Fig.~\ref{Fig:FF_GS}(c) shows that for a fixed system size, the error decreases exponentially with distance from the boundary.

We note that even though we picked representative states by filling single-particle orbitals with Fermi-Dirac probabilities at $T_E = 1$, our results also apply to other reasonable prescriptions for picking RS. For example, we can equally consider all states in some fixed $O(1)$ window about $\langle H_E \rangle_{T_E=1}$ and with some fixed spread in particle number.  This prescription will still give a $\sqrt{\frac{1}{L^{d-1}}}$ scaling of the error, but now with an improved  coefficient.

\subsection{$T>0$}

We repeat the analysis of the previous subsection, now starting with $\ket{AB}$ as an excited eigenstate of the hopping Hamiltonian $H$. We work at a finite physical temperature $T = 1$, and we can construct  $\ket{AB}$  by filling single-particle orbitals with Fermi-Dirac probabilities at $T = 1$ and $\mu = 0$. However, for computational ease, we prefer to start with the Gibbs state on $A \cup B$ instead of individual eigenstates. It is easy to check that selecting RS for the Gibbs state and excited eigenstates are equivalent upto an error of $O\left(1/L^d\right)$.

The entanglement entropy for such finite temperature states shows a volume law scaling $S_E \sim sL_A^d$, and $H_E$ acts as a genuine $d-$ dimensional Hamiltonian with excitations living everywhere in the bulk of $A$. This changes the scaling of various estimates in the previous section from $L^{d-1}$ to $L^d$, leading to an improved convergence. Since $H_E$ is still a free-fermion Hamiltonian, we pick RS according to FD probabilities at $T_E = 1, \mu_E = 0$ as before.

The inset in Fig.~\ref{Fig:FF_GS}(a) shows $\bra{\psi_A}\hat{\mathcal{A}}_i \ket{\psi_A} $ for 10,000 randomly picked representative states $\ket{\psi_A}$ in a system of linear dimension $L = 256, L_A = 128$. In this case, the spread in eigenstate expectation values appears equal for operators at all positions.  Boundary operators are not special, consistent with the volume law for the entanglement entropy of excited states.
Fig.~\ref{Fig:FF_GS}(b) (boxes and stars) shows the standard deviation of $\bra{\psi}\hat{\mathcal{A}}_i \ket{\psi_A} $ for sites $i$ lying deep in the bulk of $A$ and on the boundary, confirming the $\sqrt{\frac{1}{L^{d}}}$ scaling of the error in both cases. Note the improvement in the convergence of the EEVs at the boundary compared to zero-temperature case.

In summary, we have found RS $\ket{\psi_A}$ in free fermion systems that typically reproduce the EEVs of spatially local observables computed with $\rho_A$ in $A$. The typical error in replacing $\rho_A$ with $|\psi_A\rangle$ scales as $O(\sqrt{1/L_A^{d_{\rm eff}}})$, where $d_{\rm eff}$ is the effective dimensionality of $H_E$ and equals $d-1$ at $T=0$ and $d$ for $T>0$. For $T>0$, the convergence is independent of the distance from the boundary, while at $T=0$, the convergence is exponentially suppressed with the distance from the boundary. Thus, the boundary operators at $T=0$ exhibit the slowest convergence with system size $L_A$. Three aspects deserve re-emphasis. First, not \emph{all} states drawn from the FD distribution at $T_E=1$ (or from an energy window about $T_E=1$) are good RS. The scaling of error results are for \emph{typical} states drawn from such ensembles. Second, the convergence depends on the choice of ensemble for the RS, and can be optimized. Third, for this free fermion example, RS can be found only for a restriced class of few-body operators that live in \emph{position} space and are spread over all conserved quantities. 

Before moving on to more generic examples, let us briefly consider the implications of our free-fermion study for disordered, localized entanglement Hamiltonians that also fail to satisfy ETH. If $H_E$ is non-interacting and Anderson localized\cite{Anderson:1958ly}, its eigenstates are localized in position space. Analogous to the free-fermion example, we now expect few-body operators in a suitably defined ``momentum'' space to have an RS description\footnote{Translation invariance is broken by disorder. By ``momentum'' we just mean a set of variables obtained by an appropriate Fourier transform of the position coordinates}. Many-body localized $H_E$ deserve further thought, but here again we might expect to find RS for observables that are spread over the local integrals of motion\cite{Huse:2013kq,Serbyn:2013rt} of $H_E$.

\section{Generic eigenstates}
The previous section provided a transparent illustration of representative states for the case where $|AB\rangle$ is a Slater determinant eigenstate of a free fermion Hamiltonian. Now we turn to eigenstates of more generic, local quantum Hamiltonians which will not be Slater determinants. For such states, we expect $H_E$ to be non-integrable and we can bring the full machinery of quantum statistical mechanics and ETH to bear on our RS description. This has three important consequences:

\begin{enumerate}
\item Representative states can be used to reproduce expectation values of a much wider class of few-body operators. Unlike the free fermion case, we are no longer restricted to operators orthogonal to conserved quantities.

\item Fluctuations in EEVs for states that are close in energy are exponentially suppressed as $O(e^{-L_A^{d_{\rm eff}}})$, where $d_{\rm eff} = d \; (\mbox{or} \;d-1)$ is the effective dimensionality of $H_E$ for states obeying the volume (or area) law for the entanglement entropy \cite{Rigol:2008bh,Beugeling:2014aa}. This is to be contrasted with the free fermion case where conserved quantities led to a much larger fluctuation of $O(\sqrt{1/L_A^{d_{\rm eff}}})$ from eigenstate to eigenstate.

\item The total error in replacing $\rho_A$ with $|\psi_A\rangle$ scales as $O(1/L_A^{d_{\rm eff}})$ for reasons that will be explained below. Again, this is to be compared to a larger error that scales as $O(\sqrt{1/L_A^{d_{\rm eff}}})$ for the free fermion case.

\end{enumerate}

Points 2 and 3 above warrant further elucidation. If $H_E$ satisfies the ETH, then EEVs of an operator $\hat{\mathcal{A}}$ are hypothesized to have the form \cite{Srednicki:1994dw,Srednicki:1996aa}:
\begin{align}
\label{Eq:ETHassumption}
\bra{n}  \hat{\mathcal{A}} \ket{n} = \mathcal{A}(E) + e^{-S(E)/2} f(E) R_n
\end{align}
where $\ket{n}$ are eigenstates of $H_E$ with entanglement energy eigenvalue $E$ and $S(E)$ is the entropy (computed using $H_E$) at $E$.  Here, $\mathcal{A}(E) , f(E)$ are smooth functions of $E$ and $R_n$ is a random sign.
Since $S(E) \sim s L_A^{d_{\rm eff}}$, Eq.~\eqref{Eq:ETHassumption} implies that the dominant contribution to the EEVs comes from $\mathcal{A}(E)$. Thus, the EEVs vary smoothly with energy between neighboring eigenstates and fluctuations between eigenstates ($\sim e^{-S/2}$) are exponentially suppressed, which is the content of point 2.  Eq.~\eqref{Eq:ETHassumption} is the fundamental assumption of ETH, and the steady state properties under unitary evolution by $H_E$ and the emergence of statistical mechanics as the correct equilibrium description follow from it.

Turning now to point 3, observe that
\begin{align*}
\avg{\hat{\mathcal{A}}}_{T_E = 1} &= \frac{\textrm{Tr}\, \mathcal{A}e^{-H_E}}{\textrm{Tr}\,e^{-H_E}}  \\
&= \frac{\int dE \; e^{S(E)-E} \mathcal{A}(E)}{\int dE \; e^{S(E)-E} } + O(e^{-S/2})
\end{align*}
where the integral is over the entanglement energies. For $d>1$ and $d_{\rm eff}>0$,    $S(E)$ and $E$ are extensive in $L_A$. Thus, the integrals can be evaluated by steepest descent and expanding about the saddle point gives

\begin{align}
\avg{\hat{\mathcal{A}}}_{T_E = 1} = \mathcal{A}(\avg{E}) + O\left(\frac{1}{L_A^{d_{\rm eff}}}\right)
\label{Eq:ErrCanMC}
\end{align}
where $\avg{E} = \avg{H_E}_{T_E = 1}$ is the mean entanglement energy.

Let us now put together the various ingredients. First, a reasonable, operator independent prescription for picking representative states involves drawing eigenstates of $H_E$ with some probability in an energy window $\Delta E$ about $\avg{E}$. For example, $\Delta E \sim \sqrt{L_A^{d_{\rm eff}}}$ if states are drawn with canonical probabilities, or we can equally well pick a fixed $O(1)$ energy window. If $\mathcal{A}(E)$ varies systematically with $E$, then
\begin{align}
\mathcal{A}(E) \simeq \mathcal{A}(\avg{E}) + \frac{d\mathcal{A}}{dE}\left(\frac{\Delta E}{L_A^{d_{\rm eff}}}\right)
\label{Eq:ErrSystematic}
\end{align}
for energies within $\Delta E$ of $\avg{E}$, and we have been careful to include the fact that we're interested in local operators that depend on the energy \textit{density}. To optimize the error in the RS, let's specify an $O(1)$ energy window so the second term in Eq.\eqref{Eq:ErrSystematic} scales as $O(1/L_A^{d_{\rm eff}})$. Then, from Eqs.~\eqref{Eq:ETHassumption}, \eqref{Eq:ErrCanMC} and \eqref{Eq:ErrSystematic}, we get that
\begin{align}
\avg{\hat{\mathcal{A}}}_{T_E = 1}= \bra{n}  \hat{\mathcal{A}} \ket{n} + O\left(\frac{1}{L_A^{d_{\rm eff}}}\right)
\label{Eq:ErrRSCan}
\end{align}
when $\ket{n}$ are eigenstates of $H_E$ lying within $\Delta E $ of $\avg{E}$. This is the statement of point 3 with $\ket{n}$ acting as the representative states $|\psi_A\rangle$.\footnote{ One can
improve matters for a single operator by carefully selecting an RS
which reproduces its exact expectation value to higher accuracy but
not for the full set we wish to reproduce.}

As in the free-fermion case, we would like to support our claims with numerical evidence for some example cases. Proceeding as before would require numerically obtaining eigenstates of generic, interacting Hamiltonians which is severely limited by system size. Instead, our strategy will be to obtain $H_E$ for a particular example wavefunction and present evidence of its non-integrability by examining its level statistics. This provides strong, albeit indirect, evidence since our result, Eq.~\eqref{Eq:ErrRSCan}, follows more or less axiomatically from non-integrability and ETH.

To this end, consider the Rokhsar-Kivelson (RK) Ising wavefunction \cite{Rokhsar:1988kx},
\begin{align}
\ket{AB} = \sum_{\sigma} e^{-E_{cl}/2} \ket{\vec{\sigma}} \label{Eq:RK2d},
\end{align}
where $E_{cl}$ defines the classical anisotropic Ising model for spins $\sigma^z_{i,j} = \pm 1$ on sites $(i,j)$ of a 2D square lattice
\begin{equation}
-E_{cl} (\vec{\sigma}) = \sum_{i,j} \beta_x (\sigma^z_{i,j} \sigma^z_{i,j+1}) +\beta_y (\sigma^z_{i,j} \sigma^z_{i+1,j}) \label{Eq:Eclassical}.
\end{equation}
The probability of a given configuration is $e^{-E_{cl}(\vec{\sigma})}$. Thus, the quantum RK wavefunction reproduces classical probabilities in the $z$-basis. The RK wavefunction is the ground state of a \textit{local} Ising-symmetric parent Hamiltonian $H_{RK}(\beta_x, \beta_y)$, which is quantum critical on the same critical line as the classical 2D Ising model \cite{Henley:2004ve,Castelnovo:2005zr,Ardonne:2004ly}: $\sinh(2 \beta_x^c) \sinh(2 \beta_y^c) = 1.$ To compute $H_E$, we place the system on a cylinder of length $L_x$ and circumference $L_y$ and trace out half the cylinder with the cut parallel to the $y$ axis. The system obeys a pefect area law and $S_E \sim sL_y$. For simplicity, we take the limit $L_x \rightarrow \infty$.
We can rewrite $\ket{AB} $ in  the more convenient form
\begin{align}
\ket{AB}  &= \sum_{\sigma_L}\sum_{\sigma_R}\sqrt{\frac{T_{\sigma_L, \sigma_R } \langle \sigma_R \ket{\lambda} \langle \lambda \ket{\sigma_L}}{\lambda^2}}\ket{{\sigma_L}} \ket{{\sigma_R}} \nonumber \\
 & \equiv \sum_{\sigma_L}\sum_{\sigma_R} M_{\sigma_L, \sigma_R} \ket{\sigma_L} \ket{\sigma_R}
\label{Eq:RKTransfer}
\end{align}
where ${\sigma_L}$ ($\sigma_R$) labels the spins in the column immediately to the left (right) of the entanglement cut in A (B), and $\ket{{\sigma_L}}$ ($\ket{{\sigma_R}}$) is the RK Ising wavefunction in A (B) with the boundary spins fixed to be $\sigma_L$ ($\sigma_R$). $T_{\sigma_i, \sigma_j}$ is the (integrable) transfer matrix of the 2D Ising model. It is $2^{L_y}$ dimensional, ``transfers'' from column to column, and the indices $\sigma_{i / j}$ label the states of the $L_y$ spins in columns $i/ j$ of the lattice.
$\lambda$ is the largest eigenvalue of $T$ with corresponding eigenvector $\ket{\lambda}$. The entanglement Hamiltonian is related to the matrix $M$ though $H_E = - \log(M^{\dagger}M)$ and the entanglement energies are obtained via a singular value decomposition of the matrix $M$.

\begin{figure}[htbp]
\includegraphics[width=8cm]{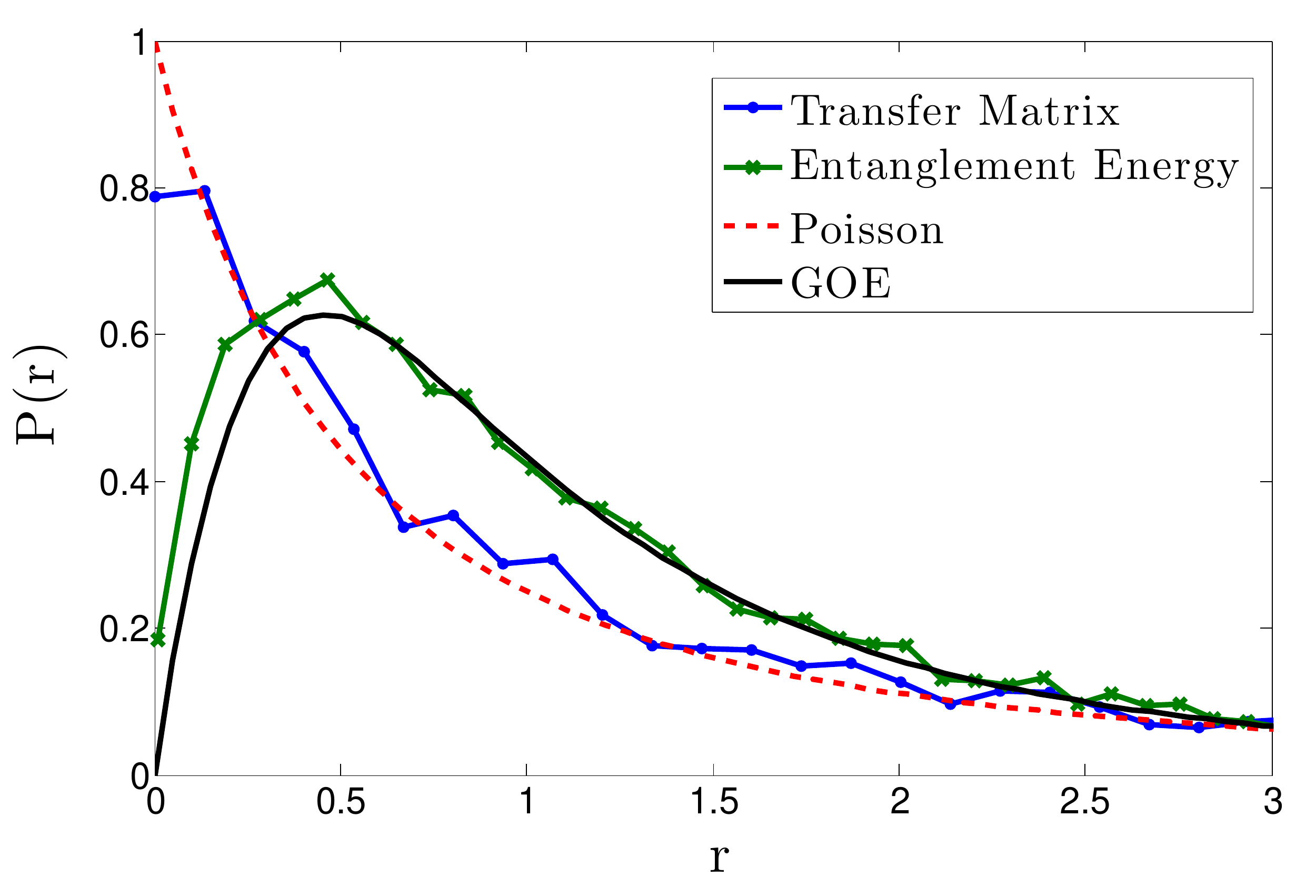}
\caption{ Level spacing ratio statistics of $H_E$ for the Rokhsar Kivelson state \eqref{Eq:RK2d} compared to the Poisson and GOE distributions. The statistics clearly look GOE consistent with a non-integrable $H_E$. This is to be contrasted with the Possoinoin statistics of the integrable transfer matrix $T_{\sigma_i, \sigma_j}$ in \eqref{Eq:RKTransfer}. $r$ refers to the ratio of subsequent level spacings, and $P(r)$ is the probability of obtaining a given $r$. The GOE form is derived in Ref.~\onlinecite{Atas:2013aa}.}
\label{Fig:RKLevel}
\end{figure}

Fig. \ref{Fig:RKLevel} shows the statistics of the ratio of adjacent level spacings of the transfer matrix $T_{\sigma_i, \sigma_j}$,  and the entanglement Hamiltonian for a paramagnetic system of size $L_y = 16$ and with $\beta_x = \beta_y = 0.43$ 
\footnote{ The entanglement Hamiltonian has translation, Ising and inversion symmetry.  We break translation symmetry by using open boundary conditions, and take the even sector with respect to both Ising and inversion symmetries to access the largest matrix size for level spacing statistics. The statistics are the same for each symmetry sector and do not depend on the boundary condition.}. Level spacings of integrable systems are known to show Poissonian statistics, while those of non-integrable systems show Gaussian Orthogonal Ensemble (GOE) statistics\cite{Bohigas}. The figure clearly shows that $H_E$ is non-integrable, even though it is so closely related to the integrable transfer matrix.

In general, we expect generic states to give generic, non-integrable entanglement Hamiltonians which are suspectible to the analysis of this section.

\section{Random states}
{
Another limit in which we can apply the idea of representative states is when $\ket{\psi_{AB}}$ is a randomly picked pure state with respect to the Haar measure on the Hilbert space of $A \cup B$. In this sense, one can find RS for almost all states!

For simplicity, we consider the ``random sign" states introduced in Ref.~\onlinecite{Grover:2013jk} below, although the same results also apply to states drawn from the Haar measure on the space of unit vectors in the entire Hilbert space as the reader can readily check.

Let $\ket{c_{AB}}$ represent a state in the computational basis on $A\cup B$.
In this basis, we define the set of  ``random sign" states via
\begin{align}
\ket{AB} = \frac{1}{\sqrt{\mathcal{N}_{A\cup B}}} \sum_{c_{AB}} \textrm{sgn}(c_{AB}) \ket{c_{AB}}
\end{align}
where the sgn function is a random variable that equals $\pm 1$ with equal probability over the $\mathcal{N}_{A\cup B}$ configurations in Hilbert space. We use $\mathcal{N}_L$ to denote the Hilbert space dimension of region $L$.
Hence for spin-$1/2$s, $\mathcal{N}_{A\cup B} = 2^N$, where $N$ is the total number of sites in the system, $\mathcal{N}_{A \cup B} = \mathcal{N}_A \mathcal{N}_B$, and $\ket{c_{AB}} = \ket{c_A} \ket{c_B}$.

For observables $\hat{\mathcal{C}}$ in some finite bounded region $C \subset A$  it is a straightforward application of the central limit theorem to
show that
\begin{align}
\bra{AB}\hat{ \mathcal{C} }\ket{AB} &= \langle\hat{ \mathcal{C} }\rangle_{T_E = 1} \nonumber\\
 & = \tr \rho_A \hat{\mathcal{C}}  = \tr \rho_C \hat{\mathcal{C}} \nonumber \\
&= \tr_{\infty} \hat{\mathcal{C}} + O\left( \frac{\mathcal{N}_C}{\sqrt{\mathcal{N}_{A \cup B} }} \right)
\label{Eq:RandCan}
\end{align}
where $\rho_C$ is the reduced density matrix of region $C$ and
$\tr_\infty \hat{\mathcal{C}} = \frac{1}{\mathcal{N}_{A\cup B}} \sum_{c_{AB}} \bra{c_{AB}} \hat{\mathcal{C}}\ket{c_{AB}}$ is the infinite temperature canonical average of observable $\hat{\mathcal{C}}$.
 Observe how $\langle\hat{ \mathcal{C} }\rangle$ is just $\tr_{\infty} \hat{\mathcal{C}}$ upto exponentially small corrections in the system size $L_{AB}$. 
 Hence our randomly picked states behave like infinite temperature states on the full system. Our first guess might be to use the results of the previous section on generic eigenstates to find representative states for $\ket{AB}$. However, those results do not apply here since $\rho_A \sim \mathbb{I} $ (up to exponentially small corrections in $L$) for such  random-sign states,  and $H_E=0$ is highly degenerate and non-generic.

Fortunately we can get around this problem by simply taking a
representative state on region $A$, $\ket{\psi_A}$, which is itself a random sign state.
The same considerations as above imply that in such a state
\begin{align}
\bra{\psi_A} \hat{\mathcal{C}} \ket{\psi_A} &= \tr_\infty \hat{ \mathcal{C}} + O\left( \frac{\mathcal{N}_C}{\sqrt{\mathcal{N}_A }} \right),
\label{Eq:RandRS}
\end{align}
which says that $\langle\hat{\mathcal{C}}\rangle$ in representative states is again $\tr_{\infty} \hat{\mathcal{C}}$ upto exponentially small corrections in $L_A$. Thus, the RS captures the same physics as the canonical ensemble of $H_E$ if the size of region $C$ is much smaller than that of $A$.  For a finite region $C$, the error in replacing the canonical ensemble with the RS is exponentially small in the size of $A$.

Note that unlike the previous two sections, we were able to pick RS for random sign states without taking into account the specific state $|AB\rangle$.
This is because of the particularly simple form that all observables take in these states. However, lest the reader be worried that these states are just trivial, we note that subsystems of such randomly picked states are close to maximally entangled with their environment as evidenced by the work of Page \cite{Page:1993rt}.}

\section{Concluding Remarks}
{
In this paper we have demonstrated that for few-body observables, the reduced density matrix of a subsystem $A$ entangled with a larger system can be replaced by a ``representative'' pure state on $A$ alone for three different
classes of states: low entanglement ground states of local quantum Hamiltonians, highly entangled randomly picked states, and highly excited eigenstates of local quantum Hamiltonians which interpolate between these
two limits in the amount of bipartite entanglement they exhibit.} The error in such a replacement is well controlled and quantified for these families
of states, and vanishes as the volume of $A$ approaches infinity. We have provided both numerical data and general arguments from quantum statistical mechanics and the ETH in support of this picture. Further, we expect that when $H_E$ is non-generic with respect to the ETH, the RS description should continue to hold for a limited set of observables and we have demonstrated this explicitly for free fermions.  

Future work could provide a more general account of classes of states $\ket{AB}$ that do, and do not, lend themselves to a description of this kind. Natural generalizations include applying these ideas to states $\ket{AB}$ with topological or symmetry-breaking order, and the reader can readily verify that the RS description naturally generalizes for local observables in these cases.

The ideas in this paper present an interesting hierarchical onion-like picture. We can replace a pure state on $A \cup B$ with a pure state on $A$ alone, which in turn can be replaced by a pure state on a subset $A_1 \subset A$, which itself can be replaced by a pure state on $A_2 \subset A_1$, and the process can be continued \emph{ad infinitum} in the limit that the volume of each subsystem approaches infinity.

Finally, we observe that the RS description is not entirely an exercise in the abstract. Isolated quantum systems in pure states form the starting point in the description of many physical phenomena. Isolated systems are of course an idealization since some degree of entanglement with the environment is inevitable, in which case the system is properly described by a density matrix. Our work suggests that the pure state description is still useful, with an error that vanishes as the system is made larger.

\section{Acknowledgements}
We thank C. R. Laumann for useful discussions. This work was supported by NSF Grant Numbers DMR 1006608, 1311781 and PHY-1005429
( VK and SLS), the John Templeton Foundation (SLS), and the Perimeter Institute
for Theoretical Physics (AC). HK is partially supported by the Samsung scholarship. Research at Perimeter Institute is supported by the Government of
Canada through Industry Canada and by the Province of Ontario through the Ministry of Research and Innovation.

\bibliography{master}

\end{document}